\documentclass[aps,pra,twocolumn,showpacs]{revtex4}

\usepackage[dvips]{graphicx}

\input{epsf}

\begin{document}

\title{Beyond Tonks-Girardeau: strongly correlated regime in quasi-one-dimensional Bose gases}

\author{G. E. Astrakharchik$^{(1,2)}$, J. Boronat$^{(3)}$, J. Casulleras$^{(3)}$, and S. Giorgini$^{(1)}$}
\address{$^{(1)}$Dipartimento di Fisica, Universit\`a di Trento and BEC-INFM, I-38050 Povo, Italy\\
$^{(2)}$Institute of Spectroscopy, 142190 Troitsk, Moscow region, Russia\\
$^{(3)}$Departament de F\'{\i}sica i Enginyeria Nuclear, Campus Nord B4-B5, Universitat Polit\`ecnica de Catalunya,
E-08034 Barcelona, Spain}

\date{\today}

\begin{abstract}
We consider a homogeneous 1D Bose gas with contact interactions and large attractive coupling constant. This system 
can be realized in tight waveguides by exploiting a confinement induced resonance of the effective 1D scattering 
amplitude. By using a variational {\it ansatz} for the many-body wavefunction, we show that for small densities 
the gas-like state is stable and the corresponding equation of state is well described by a gas of hard 
rods. By calculating the compressibility of the system, we provide an estimate of the critical density at which the 
gas-like state becomes unstable against cluster formation. Within the hard-rod model we calculate the one-body density 
matrix and the static structure factor of the gas. The results show that in this regime the system is more strongly 
correlated than a Tonks-Girardeau gas. The frequency of the lowest breathing mode for harmonically trapped systems is 
also discussed as a function of the interaction strength.        
\end{abstract}


\maketitle

The study of quasi-1D Bose gases in the quantum-degenerate regime has become a very active area of research. 
The role of correlations and of quantum fluctuations is greatly enhanced by the reduced dimensionality and 
1D quantum gases constitute well suited systems to study beyond mean-field effects \cite{Petrov}. Among these, 
particularly intriguing is the fermionization of a 1D Bose gas in the strongly repulsive Tonks-Girardeau (TG) 
regime, where the system behaves as if it consisted of noninteracting spinless fermions \cite{Girardeau}. The 
Bose-Fermi mapping of the TG gas is a peculiar aspect of the universal low-energy properties which are exhibited 
by bosonic and fermionic gapless 1D quantum systems and are described by the Luttinger liquid model \cite{Voit}. 
The concept of Luttinger liquid plays a central role in condensed matter physics and the prospect of a clean 
testing for its physical implications using ultracold gases confined in highly elongated traps is fascinating 
\cite{Monien,Recati}.    
    
Bosonic gases in 1D configurations have been realized experimentally. Complete freezing of the transverse degrees 
of freedom and fully 1D kinematics has been reached for systems prepared in a deep 2D optical lattice 
\cite{Moritz,Tolra}. The strongly interacting
regime has been achieved by adding a longitudinal periodic potential and the transition from a 1D superfluid
to a Mott insulator has been observed \cite{Esslinger}. A different technique to increase the strength of the
interactions, which is largely employed in both bosonic and fermionic 3D systems \cite{Feshbach} but has not 
yet been applied to 1D configurations, consists in the use of a Feshbach resonance. With this method one
can tune the effective 1D coupling constant $g_{1D}$ to essentially any value, including $\pm\infty$, by 
exploiting a confinement induced resonance \cite{Olshanii}. For large and positive values of $g_{1D}$ the 
system is a TG gas of point-like impenetrable bosons. On the contrary, if $g_{1D}$ is large and negative, we will 
show that a new gas-like regime is entered (super-Tonks) where the hard-core repulsion between particles becomes of
finite range and correlations are stronger than in the TG regime. Some consequences
of this new regime on the energetics of small systems in harmonic traps have already been pointed out in a 
preceding study \cite{US}. In this Letter we investigate using
Variational Monte Carlo (VMC) techniques the equation of state and the correlation functions of a homogeneous
1D Bose gas in the super-Tonks regime. We find that the particle-particle correlations decay faster than in the 
TG gas and that the static structure factor exhibits a pronounced peak. The momentum distribution and the structure 
factor of the gas are directly accessible in experiments by using, respectively, time-of-flight techniques and 
two-photon Bragg spectroscopy \cite{Esslinger}. The study of collective modes also provides a useful experimental 
technique to investigate the role of interactions and beyond mean-field effects \cite{Moritz}. Within a local density
approximation (LDA) for systems in harmonic traps we calculate the frequency of the lowest compressional mode as a 
function of the interaction strength in the crossover from the TG gas to the super-Tonks regime.   

We consider a 1D system of $N$ spinless bosons described by the following contact-interaction Hamiltonian
\begin{equation}
H=-\frac{\hbar^2}{2m}\sum_{i=1}^{N}\frac{\partial^2}{\partial z_i^2}+g_{1D}\sum_{i<j}\delta(z_{ij}) \;,
\label{Hamiltonian}
\end{equation} 
where $m$ is the mass of the particles, $z_{ij}=z_i-z_j$ denotes the interparticle distance between particle 
$i$ and $j$ and $g_{1D}$ is the coupling constant which we take large and negative. The study of the scattering 
problem of two particles in tight waveguides yields the following result for the effective 1D coupling constant 
$g_{1D}$ in terms of the 3D $s$-wave scattering length $a_{3D}$
\cite{Olshanii}
\begin{equation}
g_{1D}=-\frac{2\hbar^2}{ma_{1D}}=\frac{2\hbar^2a_{3D}}{ma_\perp^2}\frac{1}{1-Aa_{3D}/a_\perp} \;,
\label{g1D}
\end{equation} 
where $a_\perp=\sqrt{\hbar/m\omega_\perp}$ is the characteristic length of the transverse harmonic confinement
producing the waveguide and $A=|\zeta(1/2)|\sqrt{2}=1.0326$, with $\zeta(\cdot)$ the Riemann zeta-function. 
The confinement induced resonance is located at 
the critical value $a_{3D}^c=a_\perp/A$ and corresponds to the abrupt change of $g_{1D}$ from large positive 
values ($a_{3D}\lesssim a_{3D}^c$) to large negative values ($a_{3D}\gtrsim a_{3D}^c$). The renormalization 
(\ref{g1D}) of the effective 1D coupling constant has been recently confirmed in a many-body calculation of
Bose gases in highly elongated harmonic traps using quantum Monte Carlo techniques \cite{US}.

For positive $g_{1D}$, the Hamiltonian (\ref{Hamiltonian}) corresponds to the Lieb-Liniger (LL) model. The 
ground state and excited states of the LL Hamiltonian have been studied in detail \cite{LL} and, in particular,
the TG regime corresponds to the limit $g_{1D}=+\infty$. The ground state of the Hamiltonian (\ref{Hamiltonian}) 
with $g_{1D}<0$ has been investigated by McGuire \cite{McGuire} and one finds a soliton-like state with energy 
$E/N=-mg_{1D}^2(N^2-1)/24\hbar^2$. The lowest-lying gas-like state of the Hamiltonian (\ref{Hamiltonian}) with 
$g_{1D}<0$ corresponds to a highly-excited state that is stable if the gas parameter $na_{1D}\ll 1$, where $n$ 
is the density and $a_{1D}$ is the 1D effective scattering length defined in Eq. (\ref{g1D}). This state can be 
realized in tight waveguides by crossing adiabatically
the confinement induced resonance. The stability of the gas-like state can be understood from a simple estimate of
the energy per particle. For a contact potential the interaction energy $E_{int}/N=g_{1D} n g_2(0)/2$ can be 
written in terms of the local two-body correlation function $g_2(0)=\langle\psi^\dagger(z)\psi^\dagger(z)\psi(z)
\psi(z)\rangle/n^2$, where $\psi^\dagger$, $\psi$ are the creation and annihilation particle operators. In the limit 
$g_{1D}\to-\infty$ one can use for the correlation function the result in the TG regime~\cite{Gangardt} 
$g_2(0)\simeq\pi^2n^2a_{1D}^2/3$, which does not depend on the sign of $g_{1D}$. In the 
same limit the kinetic energy can be estimated by $E_{kin}/N\simeq\pi^2\hbar^2n^2/(6m)$, corresponding to the energy
per particle of a TG gas. For the total energy $E=E_{kin}+E_{int}$ one finds the result
$E/N\simeq \pi^2\hbar^2n^2/(6m)-\pi^2\hbar^2n^3a_{1D}/(3m)$,
holding for $na_{1D}\ll 1$. For $na_{1D}<0.25$ this equation of state yields a positive compressibility 
$mc^2=n\partial\mu/\partial n$, where $\mu=dE/dN$ is the chemical potential and $c$ is the speed of sound, corresponding
to a stable gas-like phase. We will show that a more precise estimate gives that the gas-like state is stable against 
cluster formation for $na_{1D}\lesssim 0.35$.

The analysis of the gas-like equation of state is carried out using the VMC technique. The trial wavefunction employed
in the calculation is of the form $\psi_T(z_1,...,z_N)=\prod_{i<j}f(z_{ij})$, where the two-body Jastrow term 
is chosen as
\begin{equation}
f(z)=\left\{ \begin{array}{cll}  \cos[k(|z|-\bar{Z})]& \mbox{ for }
&   |z| \le \bar{Z}   \\
1   &  \mbox{ for } & |z|>\bar{Z} \;.
\end{array}\right.
\label{trialwf}
\end{equation}
The cut-off length $\bar{Z}$ is a variational parameter, while the wave vector $k$ for a given $\bar{Z}$ is chosen
such that the boundary condition at $z=0$ imposed by the $\delta$-function potential is satisfied: $-k\tan(k\bar{Z})
=1/a_{1D}$. For distances smaller than the cut-off length, $|z|\le\bar{Z}$, the above wave function corresponds to
the exact solution with positive energy of the two-body problem with the interaction potential $g_{1D}\delta(z)$.
For $g_{1D}<0$ ($a_{1D}>0$) the wavefunction $f(z)$ changes sign at a nodal point which, for $\bar{Z}\gg a_{1D}$, 
is located at $|z|=a_{1D}$. The variational energy is calculated through the expectation value of the Hamiltonian
(\ref{Hamiltonian}) on the trial wave function $E=\langle\psi_T |H|\psi_T\rangle/\langle\psi_T|\psi_T\rangle$.    
In the calculations we have used $N=100$ particles with periodic boundary conditions and because of the negligible 
dependence of the variational energy on the 
parameter $\bar{Z}$ we have used in all simulations the value $\bar{Z}=L/2$, where $L$ is the size of the simulation box.
Calculations carried out with larger values of $N$ have shown negligible finite size effects.

\begin{figure}
\begin{center}
\includegraphics*[width=7cm]{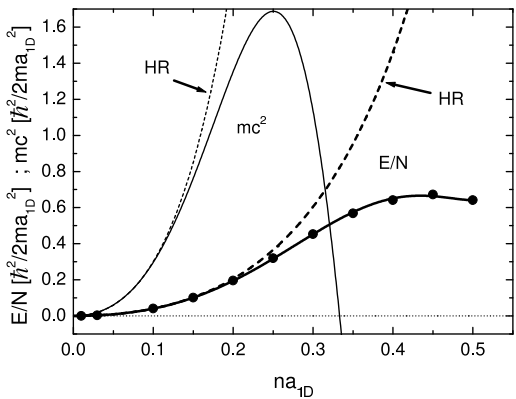}
\caption{Energy per particle and compressibility as a function of the gas parameter $na_{1D}$. Solid symbols and 
thick solid line: VMC results and polynomial best fit; thick dashed line: HR equation of state [Eq. (\ref{HR})].
Thin solid and dashed line: compressibility from the best fit to the variational equation of state and from the HR
equation of state respectively.}
\label{fig1}
\end{center}
\end{figure}

The results for the variational energy as a function of the gas parameter $na_{1D}$ are shown in Fig.~\ref{fig1} with  
solid symbols. For small 
values of the gas parameter our variational results agree very well with the equation of state of a gas of hard-rods (HR) 
of size $a_{1D}$ (thick dashed line). The HR energy per particle can be calculated exactly from the energy of a TG gas by 
accounting for the excluded volume \cite{Girardeau}
\begin{equation}
\frac{E_{HR}}{N}=\frac{\pi^2\hbar^2n^2}{6m}\frac{1}{(1-na_{1D})^2} \;.
\label{HR}
\end{equation}
For larger values of $na_{1D}$, the variational energy increases with the gas parameter more slowly than in the HR 
case and deviations are clearly visible. By fitting a polynomial function to our variational results we obtain the
best fit shown in Fig.~\ref{fig1} as a thick solid line. The compressibility obtained from the best fit is shown in 
Fig.~\ref{fig1} as a thin solid line and compared 
with $mc^2$ of a HR gas (thin dashed line). As a function of the gas parameter the compressibility shows a maximum and 
then drops abruptly to zero. The vanishing of the compressibility implies that 
the system is mechanically unstable against cluster formation. Our variational estimate yields $na_{1D}\simeq 0.35$ 
for the critical value of the density where the instability appears. This value coincides with the critical density 
for collapse calculated in the center of the trap for harmonically confined systems \cite{US}. It is worth noticing 
that the VMC estimate of the energy of the system can be extended beyond the instability point, as shown in Fig.~\ref{fig1}.
This is possible since the finite size of the simulation box hinders the long-range density fluctuations that would break
the homogeneity of the gas. This feature is analogous to the one observed in the quantum Monte-Carlo characterization of
the spinodal point in liquid $^4$He \cite{Boronat}.

As shown in Fig.~\ref{fig1}, the HR model describes accurately the equation of state for small values of the gas 
parameter. A similar accuracy is therefore expected for the correlation functions of the system. The correlation 
functions of a HR gas of size $a_{1D}$ can be calculated from the exact wave function \cite{Nagamiya}
$\psi_{HR}=\prod_{i<j}|\sin[\pi(z_i^\prime-z_j^\prime)/L]|$,
where the set of coordinates $\{z_j^\prime\}$ is obtained from the set $\{z_j\}$ with the ordering $z_1<z_2-a_{1D}
<z_3-2a_{1D}<...<z_N-(N-1)a_{1D}$ using the transformation
$z_j^\prime=z_j-ja_{1D}$, with $j=1,2,...,N$.
We calculate the static structure factor $S(k)$, which in terms of the density fluctuation operator 
$\rho_k=\sum_{i=1}^N e^{ikz_i}$, is defined as
\begin{equation}
S(k)=\frac{1}{N}\frac{\langle\psi_{HR}|\rho_k\rho_{-k}|\psi_{HR}\rangle}{\langle\psi_{HR}|\psi_{HR}\rangle} \;,
\label{Sk}
\end{equation}
and the one-body density matrix 
\begin{equation}
g_1(z)=\frac{N}{n}\frac{\int \psi_{HR}^\ast(z_1+z,..,z_N)\psi_{HR}(z_1,..,z_N)
\; dz_2 . . dz_N}{\int |\psi_{HR}(z_1,..,z_N)|^2 \;
dz_1 . . dz_N} \;.
\label{OBDM}
\end{equation}

\begin{figure}
\begin{center}
\includegraphics*[width=7cm]{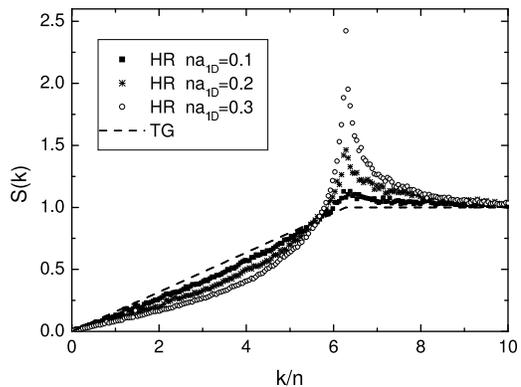}
\caption{Static structure factor $S(k)$ for a gas of HR at different values of the gas parameter $na_{1D}$ 
(symbols) and for a TG gas (dashed line).}
\label{fig2}
\end{center}
\end{figure}

\begin{figure}
\begin{center}
\includegraphics*[width=7cm]{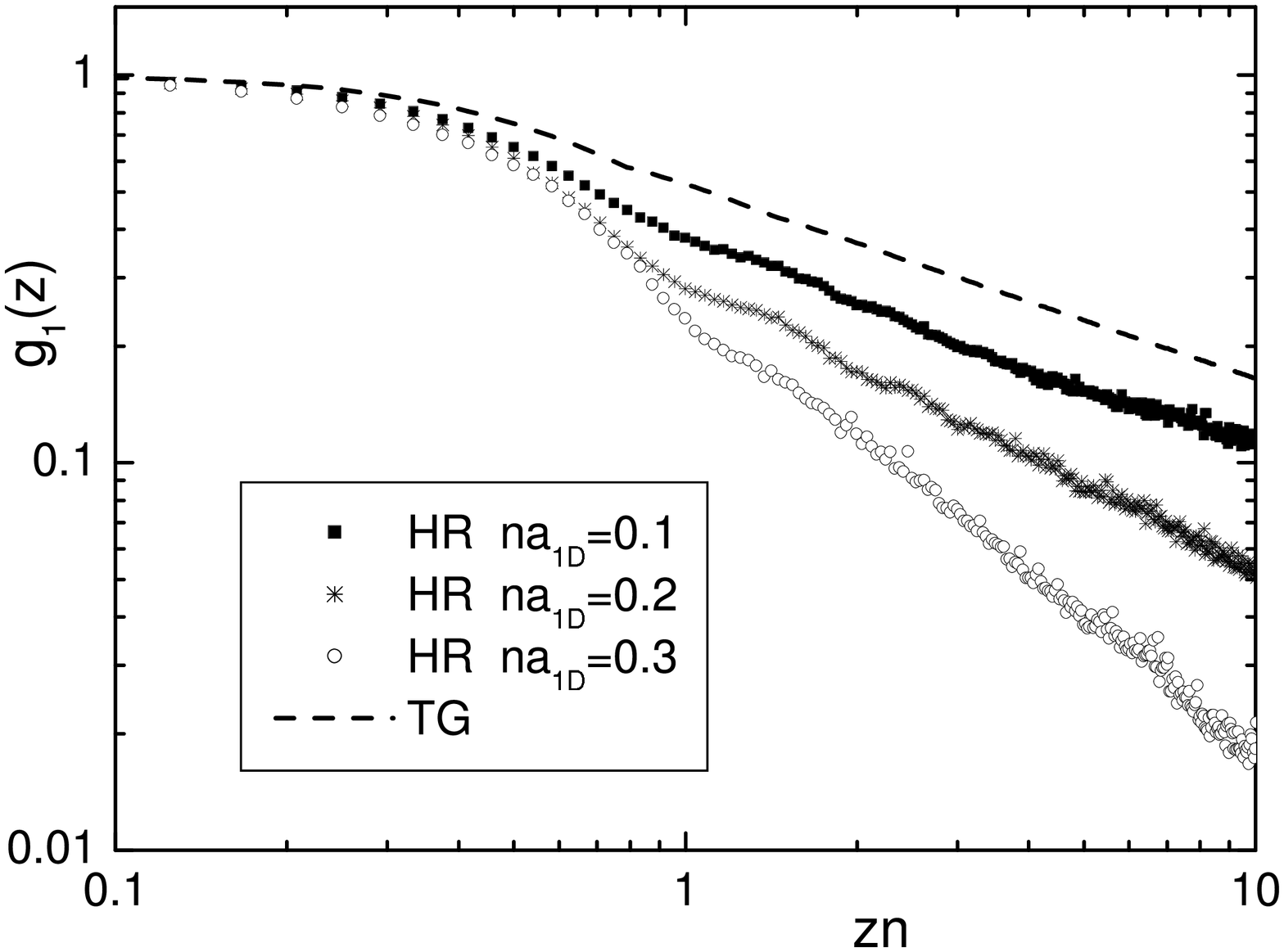}
\caption{One-body density matrix $g_1(z)$ for a gas of HR at different values of the gas parameter $na_{1D}$ 
(symbols) and for a TG gas (dashed line).}
\label{fig3}
\end{center}
\end{figure}

Contrarily to the TG case, it is not possible to obtain analytical expressions for $g_1(z)$ and $S(k)$ in the HR
problem. We have calculated them using configurations generated by a Monte Carlo simulation according to the exact 
probability distribution function $|\psi_{HR}|^2$. The results for the static structure factor are shown in 
Fig.~\ref{fig2}. Compared to $S(k)$ in the TG regime, a clear 
peak is visible for values of $k$ of the order of twice the Fermi wave vector $k_F=\pi n$ and the peak is more 
pronounced as $na_{1D}$ increases. The change of slope for small values of $k$ reflects the increase of the 
speed of sound $c$ with $na_{1D}$. The 
long-range behavior of $g_1(z)$ can be obtained from the hydrodynamic theory of low-energy excitations 
\cite{Reatto}. For $|z|\gg\xi$, where $\xi=\hbar/(\sqrt{2}mc)$ is the healing length of the system, one finds the 
following power-law decay
\begin{equation}
g_1(z)\propto1/|z|^\alpha \;,
\label{OBDM1}
\end{equation}
where the exponent $\alpha$ is given by $\alpha=mc/(2\pi\hbar n)$. For a TG gas $mc=\pi\hbar n$, and thus 
$\alpha_{TG}=1/2$. For a HR gas one finds $\alpha=\alpha_{TG}/(1-na_{1D})^2$ and thus $\alpha>\alpha_{TG}$.
This behavior at long range is clearly shown in Fig.~\ref{fig3} where we compare $g_1(z)$ of a gas of HR with 
$na_{1D}=0.1$, 0.2 and 0.3 to the result of a TG gas \cite{Jimbo}. The long-range power-law decay of $g_1(z)$ is  
reflected in the infrared divergence of the momentum distribution $n(k)\propto 1/|k|^{1-\alpha}$ holding for 
$|k|\ll 1/\xi$. A gas of HR exhibits a weaker infrared divergence compared to a TG gas.  
The correlation functions of a HR gas at $na_{1D}=0.1$, 0.2 should accurately describe the physical situation of a Bose 
gas with large and negative $g_{1D}$. For $na_{1D}=0.3$ we expect already some deviations from the HR model, as
it is evident from the equation of state in Fig.~\ref{fig1}, which should broaden the peak in $S(k)$ and decrease the 
slope of the power-law decay in $g_1(z)$ at large distances. The analysis of correlation functions clearly shows that 
the super-Tonks regime corresponds to a Luttinger liquid where short range correlations are significantly stronger than 
in the TG gas. 

\begin{figure}
\begin{center}
\includegraphics*[width=7cm]{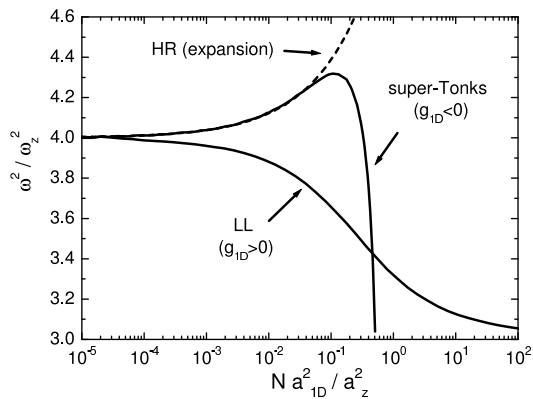}
\caption{Square of the lowest breathing mode frequency, $\omega^2$, as a function of the coupling strength 
$Na_{1D}^2/a_z^2$ for the LL Hamiltonian ($g_{1D}>0$) and in the super-Tonks regime ($g_{1D}<0$). The dashed line 
is obtained from the HR expansion (see text).}
\label{fig4}
\end{center}
\end{figure}

Another possible experimental signature of the super-Tonks regime can be provided by the study of collective modes. To this 
aim, we calculate the frequency of the lowest compressional mode of a system of $N$ particles in a harmonic potential 
$V_{ext}=\sum_{i=1}^N m\omega_z^2z_i^2/2$. We make use of LDA which allows us
to calculate the chemical potential  of the inhomogeneous system $\tilde{\mu}$ and the density profile $n(z)$ from
the local equilibrium equation $\tilde{\mu}=\mu[n(z)]+m\omega_z^2z^2/2$,
and the normalization condition $N=\int_{-R}^R n(z) dz$, where $R=\sqrt{2\tilde{\mu}/(m\omega_z^2)}$ is the size of
cloud. For densities $n$ smaller than the critical density for cluster formation, $\mu[n]$ is the equation of state of 
the homogeneous system derived from the fit to the VMC energies (Fig.~\ref{fig1}). From the knowledge of the density 
profile $n(z)$ one can 
obtain the mean square radius of the cloud $\langle z^2\rangle=\int_{R}^R n(z)z^2 dz /N$ and thus, making use of the
result \cite{Menotti}
\begin{equation}
\omega^2=-2\frac{\langle z^2\rangle}{d\langle z^2\rangle/d\omega_z^2} \;,
\label{cmode}
\end{equation}
one can calculate the frequency $\omega$ of the lowest breathing mode. Within LDA, the result will depend only on the 
dimensionless parameter $Na_{1D}^2/a_z^2$, where $a_z=\sqrt{\hbar/m\omega_z}$ is the harmonic oscillator length. 
For $g_{1D}>0$, {\it i.e.} in the case of the LL Hamiltonian, the frequency of the lowest compressional mode increases 
from $\omega=\sqrt{3}\omega_z$ in the weak-coupling mean-field regime ($Na_{1D}^2/a_z^2\gg 1$) to $\omega=2\omega_z$ 
in the strong-coupling TG regime ($Na_{1D}^2/a_z^2\ll 1$). The results for
$\omega$ in the super-Tonks regime are shown in Fig.~\ref{fig4} as a function of the coupling strength. In the regime
$Na_{1D}^2/a_z^2\ll 1$, where the HR model is appropriate, we can calculate analytically the first correction to the 
frequency of a TG gas. One finds the result $\omega=2\omega_z [1+(16\sqrt{2}/15\pi^2)(N a_{1D}^2/a_z^2)^{1/2}+...]$.
Fig.~\ref{fig4} shows that this expansion accurately describes the  frequency of the breathing mode when 
$Na_{1D}^2/a_z^2\ll 1$, for larger values of the coupling strength the frequency reaches a maximum and drops to zero
at $Na_{1D}^2/a_z^2\simeq 0.6$. The observation of a breathing mode with a frequency larger than $2\omega_z$ would be 
a clear signature of the super-Tonks regime.

In conclusion we have pointed out the existence of a strongly correlated regime in quasi-1D Bose gases beyond the
Tonks-Giradeau regime. This regime can be entered by exploiting a confinement induced resonance of the effective 1D
scattering amplitude. We calculate the equation of state of the gas in the super-Tonks regime using VMC and we estimate
the critical density for the onset of instability against cluster formation. The static structure factor and one-body
density matrix are calculated exactly within the hard-rod model, which provides the correct description of the system
for small values of the gas parameter. For harmonically trapped systems we provide explicit predictions for the 
frequency of the lowest compressional mode.

Acknowledgements: Useful discussions with D. Blume are gratefully acknowledged. GEA and SG acknowledge support by the 
Ministero dell'Istruzione, dell'Universit\'a e della Ricerca (MIUR). JB and JC acknowledge support from DGI (Spain) 
Grant No. BFM2002-00466 and Generalitat de Catalunya Grant No. 2001SGR-00222.

\end{document}